\documentclass[9pt,article,twoside]{rmaa-rho-class/rmaa-rho}
\RMxAAtemplatetype{\RMxAA} 
\usepackage{array}

\newcolumntype{P}[1]{>{\centering\arraybackslash}p{#1}}

\vol{62}
\pages{110-132}
\thisyear{2026}
\doi{\href{https://doi.org/10.22201/ia.30618649e.2026.62.02.7465}{https://doi.org/10.22201/ia.30618649e.2026.62.02.7465}}

\title{High-resolution spectroscopy of the bipolar proto-planetary nebula Hen 3-1475, the Garden Sprinkler Nebula}

\author[1,$\dagger$]{Ana Valeria Beltrán-Sánchez}
\author[1,$\dagger$]{Miriam Peña \orcidlink{0009-0007-5891-420X}}
\author[2]{Mudumba Parthasarathy}

\affil[1]{Universidad Nacional Autónoma de México, Instituto de Astronomía, Ciudad Universitaria, Cd. de México, México}
\affil[2]{Indian Institute of Astrophysics, Bangalore 560034, India}

\leadauthor{Beltrán-Sánchez et al.}
\smalltitle{Proto-planetary nebula Hen 3-1475}

\corres{Ana Valeria Beltrán-Sánchez}
\email{avbeltran@astro.unam.mx}

\received{}
\accepted{}

\license{Texto de la licencia aquí}

\setbool{rho-abstract}{true} 
\setbool{rho-resumen}{true} 

\begin{abstract}
  A detailed analysis of high-resolution spectra in the optical and near-infrared range of the central zone of the proto-planetary nebula Hen\,3-1475, obtained in 2006 and 2024, is conducted. The spectrum primarily revealed stellar emission of H Balmer and Paschen series, lines of iron (\ion{Fe}{I}, \ion{Fe}{II}, and {[}\ion{Fe}{II}{]}), and other significant lines of \ion{He}{I}, \ion{Ca}{II}, {[}\ion{Ca}{II}{]}, and \ion{O}{I}.  
 Several multiplets of \ion{Fe}{II}, the H Balmer series, \ion{Ca}{II} H and K, \ion{Na}{I} D and \ion{He}{I} lines exhibit \textit{P-Cygni} profiles and bump-like features, which are clear indicators of strong stellar winds. The terminal velocity of the winds was calculated, reaching values up to $\sim 800$ km s$^{-1}$.  
 {[}\ion{Fe}{II}{]} and \ion{Fe}{I} lines  were used as gas density indicators. 
In these observations, no photoionized nebular lines of planetary nebulae were identified, suggesting that the central star has not reached the temperature required to photoionize the nebula. A comparison between the observations from 2006 and 2024 reveals some few significant differences, being the most important the probably ejection of a new jet emerging from the central zone and the detection of a broad component adjacent to \ion{H}{$\alpha$}, probably originating from emission associated with a jet at 700 km s$^{-1}$.
High-spectral resolution observations of the knot NW1 obtained on August 2024, are also presented. This knot shows a typical spectrum of shocked gas. The emission of this knot has changed, presently not showing [\ion{O}{III}] 5007 and other ionized lines that were  reported in 2006. 
\end{abstract}


\begin{resumen}
    Se presenta un análisis detallado de espectros de alta resolución en el rango óptico e infrarrojo cercano de la zona central de la proto nebulosa planetaria Hen\,3-1475, obtenidos en 2006 y 2024. El espectro muestra principalmente líneas de emisión estelares de las series de Balmer y Paschen del hidrógeno, líneas de hierro (\ion{Fe}{I}, \ion{Fe}{II} y {[}\ion{Fe}{II}{]}), así como otras líneas relevantes de \ion{He}{I}, \ion{Ca}{II}, {[}\ion{Ca}{II}{]} y \ion{O}{I}. Varios multipletes de \ion{Fe}{II}, las líneas de la serie de Balmer del hidrógeno, \ion{Ca}{II} H and K, \ion{Na}{I} D y \ion{He}{I} presentan perfiles tipo \textit{P-Cygni} y estructuras en forma de “abultamiento”, que son indicadores claros de la presencia de vientos estelares intensos. La velocidad terminal de estos vientos se estimó en valores de hasta $\sim 800$ km s$^{-1}$. Las líneas de {[}\ion{Fe}{II}{]} y \ion{Fe}{I} se utilizaron como indicadores de la densidad del gas.
En estas observaciones no se identificaron líneas nebulares fotoionizadas típicas de nebulosas planetarias, lo que sugiere que la estrella central aún no alcanza la temperatura necesaria para fotoionizar la nebulosa. La comparación entre las observaciones de 2006 y 2024 revela algunas diferencias significativas, siendo la más importante la probable eyección de un nuevo chorro emergente de la zona central y la detección de un componente ancho adyacente a \ion{H}{$\alpha$}, probablemente originado por emisión asociada a un chorro a 700 km s$^{-1}$.
También se presentan observaciones de alta resolución espectral del nudo NW1 obtenidas en agosto de 2024, cuyo espectro es característico de gas chocado. La emisión de este nudo ha cambiado con respecto a observaciones previas, pues actualmente no muestra la línea [\ion{O}{III}] 5007 ni otras líneas de gas ionizado que se habían reportado en 2006.
\end{resumen}

\begin{document}

\maketitle
\pagestyle{fancy}\thispagestyle{firststyle}


\section{INTRODUCTION}

  Proto-planetary nebulae (pPNe) are objects in transition from the asymptotic giant branch (AGB) to the planetary nebula stage (PNe). During the AGB phase, the outer layers of the star are expelled with mass loss rates of  $10^{-8}$--$10^{-4} $M$_{\odot}$ yr$^{-1}$, forming circumstellar envelopes \citep{2005Herwig}. It is during this transition that the wide variety of morphologies emerges. Although substantial progress has been made, many of the detailed shaping mechanisms are still not fully understood. During the pPNe phase, these objects generate very fast winds that interact violently with the surrounding gas, resulting in changes in their morphology and kinematics \citep{2003ApJ...599L..87S}.

Hen\,3-1475 (PN G009.3+05.7, IRAS\,17423-1755, also known as the "Garden Sprinkler Nebula") was initially classified as a B-type emission-line star \citep{Henize1976}. It was thought to be a Herbig Be star, due to its strong \ion{H}{I}, \ion{O}{I}, \ion{Fe}{II}, and \ion{Ca}{II} lines, as well as its \textit{P-Cygni} profiles. However, \citet{Partha89} were the first to determine that it is actually a transitioning object. They identified Hen\,3-1475 as an IRAS source and, based on the analysis of its IRAS data and circumstellar dust shell parameters, they classified it as a hot post-AGB star and concluded that it may be evolving rapidly towards the young PN stage. 

Later \citet{Riera95} showed that Hen\,3-1475 exhibits N-enrichment and far-infrared colors similar to those of known planetary nebulae and corroborated the \textit{P-Cygni} Balmer line profiles of the central star, indicating mass-loss and, furthermore, high-velocity mass outflows. With subsequent observations \citet{2001ApJ...553L.173S} suggested two different high-velocity winds related to two blue-shifted absorption features in H$\alpha$.

 Observations by \citet{2000Partha} of this object revealed that the Balmer lines showing \textit{P-Cygni}  are blue-shifted by about $-400$ km s$^{-1}$, corroborating the mass loss from the post-AGB central star. 

Different distances have been derived for Hen\,3-1475, from 2.5 kpc \citep{Partha89} to 8.3 kpc \citep{2001Borkowski}.  Hen\,3-1475 was observed by GAIA; in the DR3 catalogue, it appears with the number Gaia DR3 4120637086125583360 showing a measured parallax of $0.3145\pm 0.3048$ mas, and a G magnitude of $G=12.237\,\pm \,0.007$. 
Based on the GAIA parallax data \citet{2021Bailer} derived a geometric distance of 5.32 kpc with a large uncertainty. This value is consistent, within uncertainties, with the value $5.8\pm 0.9$ kpc, originally determined by \citet{2003Riera}. Riera's estimated distance will be used in the following.  

This object possesses a highly collimated bipolar structure with an S-shaped string of point-symmetric [\ion{N}{ii}] bright knots extending over approximately 17'' along the main axis \citep{Xuan}. {bf This size corresponds to 0.48 pc at a distance of 5.8 pc.} 

High-resolution spectroscopy shows high-velocity jets at 1200 km s$^{-1}$  \citep{2001Borkowski} and ultra fast winds up to 2300 km s$^{-1}$ \citep{2001ApJ...553L.173S}. Spectra of the knots have been partially reproduced using simple bow shock models with expanding velocities of $\sim$150 to 200 km s$^{-1}$ \citep{2006Riera}. 
\citet{2004A&A...419..991V} presented 2D and 3D gasdynamical simulations suggesting a time-dependent jet model with an ejection velocity history composed of a sinusoidal mode superimposed on a linear increase of the ejection velocity with time.

OH maser emission was detected by \citet{1991A&A...248..209T} with the main line at 1667 MHz. \citet{2001MNRAS.322..280Z} have suggested that this emission arises from a disk or shell of neutral gas expanding at 25 km s$^{-1}$. In support of the idea of large amounts of neutral gas, different authors, including \citet{1995Knapp} and \citet{2001Bujarrabal}, have observed a CO-expanding torus. 

By analyzing the IR colors of the object, \citet{2003Rodrigues} derived an interstellar reddening of $A(V)\sim$3.2 mag, larger than the value of 2.0$\pm$0.4 derived by \citet{Riera95}. Additionally, due to circumstellar material around the central star, \citet{2003Rodrigues} calculated a circumstellar extinction of $A(V)\approx 9.0$ mag.

 Because of its jet formation/collimation, \citet{Xuan} have suggested that the central object is possibly a binary star. Several other authors have also proposed this, based on different nebular characteristics, which will be discussed in more detail in \S\ref{sec:binary}.

 X-ray emission was detected by \citet{2003ApJ...599L..87S} coming mostly from the brightest knot identified as NW1 in Fig.1 by \citet{2003Riera}. 

\begin{figure}
\centering
  \includegraphics[width=0.98\linewidth]{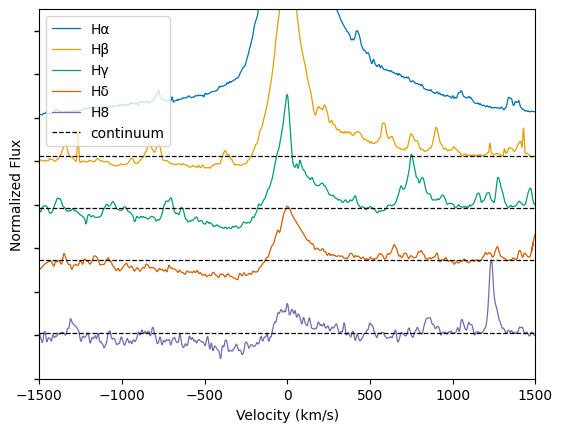}
  \caption{Balmer series lines (\ion{H}{$\alpha$}, \ion{H}{$\beta$}, \ion{H}{$\gamma$}, \ion{H}{$\delta$} and H\,8) from 2006 observations. \ion{H}{$\epsilon$} is missing due to blending with a calcium line. The black dotted lines indicate the continuum making clear the \textit{P-Cygni} profile.}
  \label{fig:balmerlines}
\end{figure}

\citet{2011Manteiga} indicated that the central star had possibly already become hot enough to ionize the surrounding gas, as they detected the nebular emission line [\ion{Ne}{ii}] at 12$\mu$m. In their paper and in \citet{2001Borkowski} it is suggested that Hen\,3-1475 is probably a relatively massive ($M>3M_{\odot}$) O-rich post-AGB star, as they found an absorption band at 3.1 $\mu m$ identified as water ice.
In a recent work by \citet{khouri2025} the $^{17}$O$/^{18}$O isotopic ratio is used to estimate the initial mass of Hen\,3-1475, finding $M<2M_{\odot}$, a lower result than in the previous works. It is mentioned that despite this initial mass, Hen\,3-1475 continues to be an O-rich source. 

In this work, we present a detailed spectroscopic analysis of this proto planetary nebula (pPN) in two epochs, with the purpose of complementing the optical information using high-resolution spectra to give some insight on the central star and the nebula.  In particular, we are interested in determining if the central star has already become hot enough to photoionize the nebula. Besides, we aim to contribute to the discussion on the possible binarity of the central star. We also search for significant changes in the stellar and nebular emission of this object in the past 20 years in order to analyze its temporal evolution on a short time scale.

This paper is organized as follows, in \S\ref{sec:obs} the two sets of observations, discussed in this work, are presented;
in \S\ref{sec:ferosobs}, observations from 2006 obtained with the FEROS spectrograph are described. In \S\ref{sec:newobs}  details on the 2024 observations obtained at the Observatorio Astron\'omico Nacional San Pedro M\'artir, México (OAN-SPM), and data reduction procedures for the Echelle REOSC spectrograph data are provided. \S\ref{sec:feroscentral} elaborates on the spectral lines found in the spectra, describing how they were measured, and what their profiles and velocities are. This section is further subdivided to delve into the detailed characteristics of different spectral lines. In \S\ref{sec:binary} the possible binary nature of this object is discussed, incorporating findings from previous studies, as well as those presented in this work.  \S\ref{sec:analysis2024} outlines the differences found between the 2006 and 2024 observations. In \S\ref{sec:NW1} the characteristics of knot NW1 are described. Our conclusions are presented in \S\ref{sec:conclusion}. 

\section {OBSERVATIONS} \label{sec:obs}
Two sets of observations, obtained at different epochs,  are presented and analyzed in this work. The first one was obtained with the spectrograph FEROS at La Silla Observatory in Chile in 2006,  and the second one was obtained in 2024 at the OAN-SPM. These observations are described in the following.

\subsection{OBSERVATIONS FROM FEROS} \label{sec:ferosobs}

Spectrophotometric data from the central zone of Hen\,3-1475 were obtained in April 2006, at La Silla Observatory, with the Echelle spectrograph FEROS attached to the MPG/ESO 2.2-m telescope. The spectral resolution of these data is $R\sim 48 000$ and the wavelength coverage goes from $\sim 3800$ {\AA} to $\sim 9200$ {\AA}. 
FEROS, as its name stands for, has two fibers illuminated via 2 arcsec circular apertures on the sky, one for the object and one for the sky or the wavelength calibration lamp. This observation included the central star.

The total exposure time of this observation was 2700 s. The spectrum was divided into four spectra of different wavelength ranges. The first one covers the range from $\sim 3800$ to $\sim 5000$ {\AA}, the second, from $\sim 4900$ to $\sim 6400 $ {\AA}, the third, from $\sim 6300$ to $\sim 7800$ {\AA}, and the fourth one from $\sim 7800$ to $\sim 9200$ {\AA} (see Figure \ref{fig:FEROS}).  

The reduced data was obtained from the ESO archives. 
Many stellar lines are detected in the spectrum.
All present lines were measured and listed in Table \ref{tab:Hen31475}. The line IDs are discussed in the following sections.

\begin{figure*}
\centering  
\includegraphics[width=0.7\linewidth]{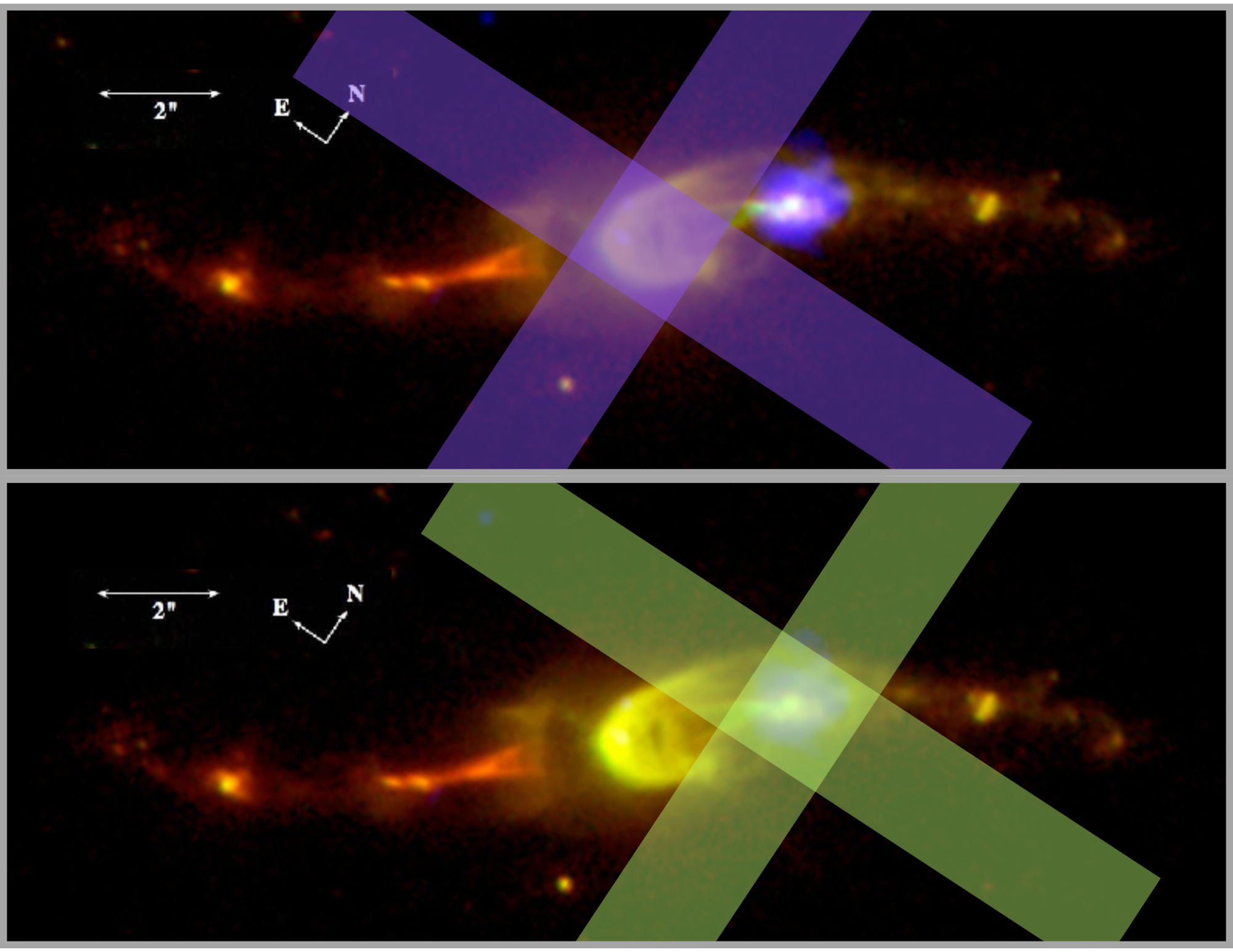}
\caption{Hubble Space Telescope (HST) F658N images of Hen\,3-1475 obtained in 1999 (red) and 2009 (green) presented by \textcolor{blue}{Fang et al. (2018)} along with OAN-SPM slit orientations for observations (performed in 2024) of the center (purple) and NW1 knot (yellow). The slits are 2 arcsec wide and 13 arcsec long, and some are oriented from E to W and others from N to S. }
  \label{fig:hst}
\end{figure*}

\subsection{OAN-SPM OBSERVATIONS AND DATA REDUCTION}\label{sec:newobs}

We conducted a new optical spectroscopic study of this object using the Echelle REOSC spectrograph attached to the 2.1-m telescope at OAN-SPM, México. In this case, the spectral resolution is $R=18000$ at $5000$ {\AA}. And the typical seeing during the observations was $\sim$ 1.5 arcsec. These observations took place from  August 27 to 30, 2024, covering a wavelength range from $\sim$ 3700 Å to $\sim$ 7000 Å. During the first 3 nights of observation, the slit was positioned passing through the center of the nebula, crossing the central star, with an E-W orientation (P.A. 270 $\deg$). On the 4th night, the slit orientation was changed to N-S (P.A. 180 $\deg$). Slit widths of 150 $\mu$ (2 arcsec) and 500 $\mu$ ($\sim$ 7 arcsec) were used, and the slit length was 13.3 arcsec. 

The  nebular knot NW1, which is the brightest knot in the S-shape structure of the nebula \citep{2003Riera}, was also observed during 3 nights. For this, the telescope was moved $\sim$2''.80 (2''-W, 2''-N) from the center (see Fig. \ref{fig:hst}), following the positions stated by \citet{Xuan}. Note that on the fourth night (August 30) the slit orientation was N-S, the same as for the data collected for the center. 
The log of observations is presented in Table \ref{tab:observations}.

Both long and short exposure times were employed to achieve good signal-to-noise ratios for faint lines and to avoid saturation of intense lines (exposure times of 1800 s, 1200 s, 600 s, 180 s, and 120 s were used). A Th-Ar lamp was taken after each scientific observation for wavelength calibration. Two standard stars from the list by \citet{1992Hamuy} were observed each night for flux calibration. The stars HR\,153, HR\,9087, and HR\,7596 were used. The slit width for the standard stars was set to 500 $\mu$ ($\sim$7 arcsec) to ensure all stellar flux was included in the slit.

\subsubsection{OAN-SPM Data Reduction}
Each 2D spectrum was cropped at the edges and bias-subtracted. Afterwards, spectra were extracted including all the emission along the slit. The sky was subtracted using parts of the slit that showed no evidence of emission, and  the extraction window was 5 arcsec. We have to note that the spectra were processed individually due to several issues encountered during the observation nights - such as guider malfunctions, which caused flux loss in some observations, and telescope oscillations, which also resulted in flux loss. Each spectrum was continuum-normalized only after this step. After the entire calibration process, spectra were combined to enhance the signal-to-noise ratio.

Wavelength calibration was performed using a Th-Ar lamp, and flux calibration was carried out with the spectra of standard stars. The fluxes of individual lines were measured using the IRAF task splot. The Full Width at Half Maximum (FWHM) was determined by fitting a Gaussian profile to the lines. 

To determine flux errors, the expression by \citet{1999Tresse}  was used:
\begin{align}
    \sigma_\text{F} = \sigma_\text{c} \text{D}\sqrt{2\text{N}_{\text{pix}}+\frac{\text{EW}}{\text{D}}}
\end{align}
where D represents the spectral dispersion in {\AA} per pix (for Echelle REOSC the value is $\sim$0.3 {\AA}/pix), \( \sigma_\text{c} \) is the mean standard deviation per pixel of the continuum measured on both sides of the line, and \( \text{N}_\text{pix} \) denotes the number of pixels spanned by the line. Thus, every value was specifically determined for each line in the spectrum. 

\begin{table}
\centering
\caption{2.1-m OAN-SPM (REOSC Echelle) observing log (August 27 corresponds to Night 1, August 28 to Night 2, and so forth).}
\resizebox{8.5cm}{!}{
\begin{tabular}{cccc}
\hline
\multicolumn{4}{|c|}{Central Star} \\ \hline
\textbf{Date}      & \textbf{Exp.} & \textbf{Spectral} & \textbf{Slit width} \\ 
\textbf{ (2024)}   &Time \textbf{(s)}& Range \textbf{(Å)}           &    \textbf{(arcsec)}                 \\ \hline\hline
August 27   & 120, 3$\times$1800      & 3700 - 7100          & 2\\ 
August 28   & 180, 2$\times$600, 1800 & 3700 - 7100          & 7\\ 
August 29   & 180, 2$\times$1800      & 3700 - 7100          & 2\\ 
August 30   & 180, 2$\times$1800      & 3700 - 7400          & 2\\ \hline
\multicolumn{4}{|c|}{NW1 knot} \\ \hline
\textbf{Date}      & \textbf{Exp. Time} & \textbf{Spectral Range} & \textbf{Slit Size} \\ 
        \textbf{ (2024)}          &        \textbf{(s)}                & \textbf{(Å)}           &    \textbf{(arcsec)}                 \\ \hline\hline
August 28   & 600, 2$\times$1800 & 3700 - 7100  & 7, 2\\
August 29   & 1800               & 3700 - 7100  & 2\\
August 30   & 1800               & 3700 - 7400  & 2\\ \hline

\end{tabular}
}
\label{tab:observations}
\end{table}

\section{Lines present in the optical spectrum}\label{sec:feroscentral}

In the spectra obtained at OAN-SPM, we generally observe the same features reported from  the FEROS observations performed nearly 20 years ago, although there are several differences that will be described later. Being the OAN-SPM observations of lower resolution, some lines appear blended; therefore, we will describe in detail the lines appearing in the FEROS data which are listed in Table \ref{tab:Hen31475}.

Observations of the central region reveal emission lines from \ion{H}{I}, \ion{He}{I}, \ion{Fe}{II},  [\ion{Fe}{II}], \ion{O}{I}, \ion{Ca}{II}, [\ion{Ca}{II}], \ion{Si}{II}, and possibly \ion{Cr}{II}, along with other ions. In total, more than 400 lines (as seen in Table \ref{tab:Hen31475}) were measured from these spectra using the IRAF task \textit{splot}. In addition, emission lines attributed to elements such as titanium, scandium,  and vanadium were identified. These lines are typically weak, and they are only detectable in high-resolution spectra.

\textit{P-Cygni} profiles are identified in the \ion{H}{I} Balmer lines, several multiplets of \ion{Fe}{II} (including \#s 27, 38, 42, 49, and 74), \ion{Si}{II} multiplet 2, \ion{Ca}{II} H and K lines, \ion{Na}{I} D doublet and \ion{He}{I} lines. These profiles indicate the presence of high-velocity winds, and the terminal velocity can be determined from the extent of the absorption component.  
In this work, the terminal velocity was calculated using the wavelength corresponding to this maximum velocity. However, since the exact rest wavelength is unclear due to contributions from different parts of the object to the emission, the systemic velocity was used as a reference. This systemic velocity was calculated as the average value determined from emission lines without \textit{P-Cygni} profiles, giving a result of $v_{\text{LSR}}\approx 39.34 \, \pm$ 0.38 km s$^{-1}$.  

It is important to consider that the uncertainty associated with the terminal velocity is significant, as many lines in the spectrum often overlap with the absorption component. 

The terminal velocities for the ions mentioned above are presented in Table \ref{tab:velHen31475}, showing a range between $\sim$200 and 800 km s$^{-1}$, indicating that different lines are formed in different zones of the wind. 

\begin{figure}
\centering
  \includegraphics[width=0.97\linewidth]{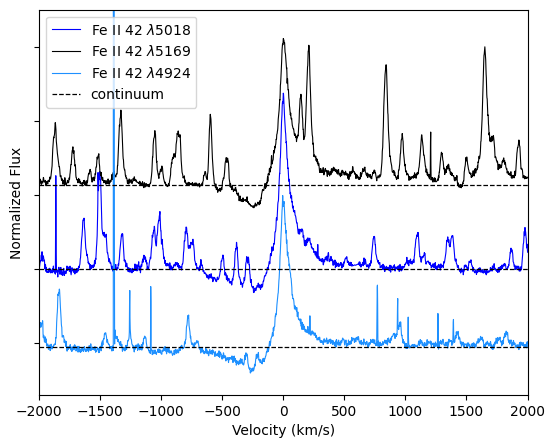}
  \caption{\ion{Fe}{II} from multiplet 42.}
  \label{fig:feii42}
\end{figure}

\subsection{Balmer lines}
Notably, the Balmer lines exhibit differences  in their profiles. For instance, \ion{H}{$\alpha$}  shows an asymmetric emission with no clear absorption but a prominent red wing, as shown in Figure \ref{fig:balmerlines}. The red wing appears in all the Balmer lines. 

As said above \textit{P-Cygni} can be detected in all the Balmer lines except \ion{H}{$\alpha$}. They are shown in Fig. \ref{fig:balmerlines}. From the blue absorption, terminal velocities are calculated and reported in Table \ref{tab:velHen31475}. 

According to \citet{2001ApJ...553L.173S}, in the spectra of the central zone \ion{H}{$\alpha$} exhibited  two blue-shifted absorption features, as observed with the HST using the Space Telescope Imaging Spectrograph (STIS) in June 1999. 
In a subsequent study by the same authors \citep[ground-based observations performed in 2003]{2008Sanchez}, \ion{H}{$\alpha$} was observed to exhibit not the two absorption components but only one. The authors attributed this to the apparent smoothing of spectra in ground-based observations, probably caused by blending of emission components from different nebular regions superimposed within the PSF. 

\subsection{The Fe lines}

As previously mentioned, the \textit{P-Cygni} profile is present in lines of several \ion{Fe}{II} multiplets, including  \#s 27, 38, 42, 48, 49, and 74. Multiplet 42 is the most prominent among these, and its lines are shown in Figure \ref{fig:feii42}, where terminal velocities derived from the blue component reach nearly 700 km s$^{-1}$.

It is worth noting that additional multiplets are observed besides those exhibiting \textit{P-Cygni} profiles. From each of these, at least two emission lines are detected. These multiplets include \#s 3, 25, 28, 30, 32, 35, 36, 37, 40, 43, 46, 55, 57, 72, 73, 182, 199, and 200.

In addition to the \ion{Fe}{II} lines, lines of various multiplets of \ion{Fe}{I} are also identified, such as \#12, 13, 15, 16, 36, 60, 62, 66, 111, 168, 169, 207, 401, 816, 1051, 1136, and 1154. These lines do not present \textit{P-Cygni} profile. Similarly, several multiplets of [\ion{Fe}{II}] are observed, including \#s 6F, 7F, 13F, 14F, 17F, 18F, 19F, 20F, 21F, and 30F, among others. These lines do not present \textit{P-Cygni} profile either.

It should be noted that the fact that we observe the latter lines means that the gas in the zone has a density limit lower than the critical density of [\ion{Fe}{ii}] which is $\sim 10^{4}-10^{5}$ cm$^{-3}$ \citep{2002Nisini}.

\begin{figure*}[ht]
\centering

\begin{subfigure}[b]{0.48\linewidth}
   \centering
   \includegraphics[width=\linewidth]{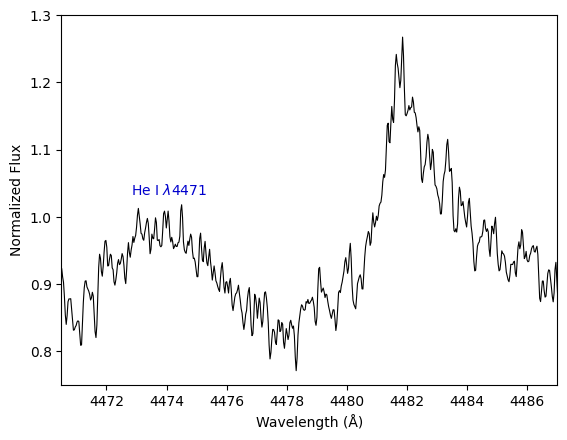}
   \caption{}
   \label{fig:hei4471}
\end{subfigure}
\hfill
\begin{subfigure}[b]{0.50\linewidth}
   \centering
   \includegraphics[width=\linewidth]{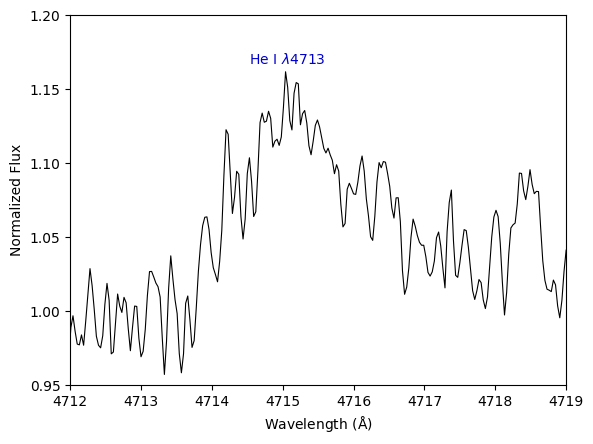}
   \caption{}
   \label{fig:hei4713}
\end{subfigure}
\hfill
\begin{subfigure}[b]{0.49\linewidth}
   \centering
   \includegraphics[width=\linewidth]{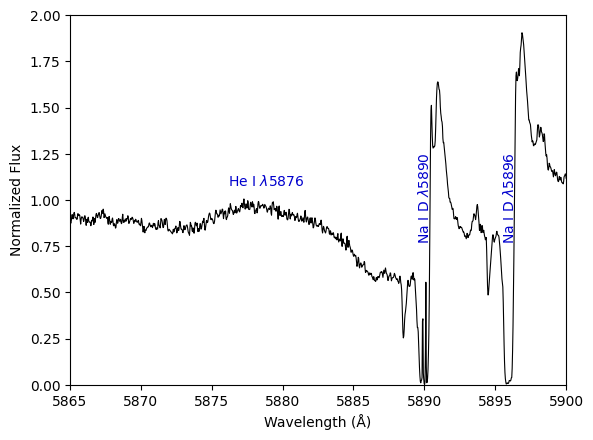}
   \caption{}
   \label{fig:sodiumlines}
\end{subfigure}
\hfill
\begin{subfigure}[b]{0.48\linewidth}
   \centering
   \includegraphics[width=\linewidth]{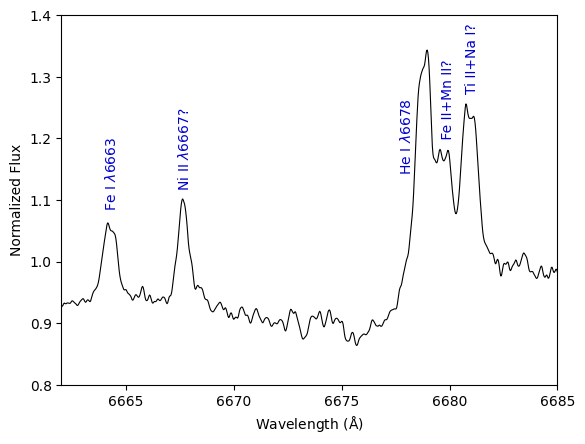}
   \caption{}
   \label{fig:6678}
\end{subfigure}
\caption{(a) \ion{He}{I} $\lambda$4471 bump and an unidentified line with \textit{P-Cygni} profile, most likely corresponding to \ion{Mg}{II}.(b) Bump + \textit{P-Cygni} profile in the \ion{He}{I} $\lambda$4713 line. (c) \ion{He}{I} $\lambda$5876 bump and the \ion{Na}{I} D lines showing \textit{P-Cygni} profile. (d) \textit{P-Cygni} profile observed in the \ion{He}{I} $\lambda$6678 line. }
\label{fig:he_na_lines}

\end{figure*}
\subsection{The He\,I and Na\,I lines}
When searching for the \ion{He}{I} $\lambda$5876  line, a noticeable bump is observed spanning over 10 {\AA} around 500 km s$^{-1}$), characterized by a non-Gaussian profile. This \ion{He}{I} line is shown in Fig.  \ref{fig:sodiumlines} together with the \ion{Na}{I} D doublet at $\lambda\lambda$5890,5896. Both Na lines present  \textit{P-Cygni} profile and their presence may indicate the existence of neutral gas around Hen\,3-1475 (this is corroborated by the potassium line at $\lambda7699$ observed in the FEROS spectrum)

In the work by \citet{Riera95}, \ion{He}{I} $\lambda$5876 is described as an absorption line; however, it is important to note that their analysis was based on a low-resolution spectrum where this line appears to be blended with the sodium lines. It is clear from Fig. \ref{fig:sodiumlines} that this blending does not occur in our observations.

The same broad feature seen in $\lambda$5876 is found in \ion{He}{I} $ \lambda$4471 (see Fig.  \ref{fig:hei4471}). 
Broad stellar emission lines are often observed in stars with strong winds. In this case, it is not clear whether the bumps in \ion{He}{I} lines have their origin in the strong wind or if they might have a different origin, as will be discussed in the following text.  

Additional helium lines are detected at 4715 {\AA}, 6678 {\AA}, and 7065 {\AA}. The $\lambda$6678 line exhibits a \textit{P-Cygni} profile, while $\lambda\lambda$4715 and 7065 show a broad emission, similar to the ones mentioned earlier, and a slight \textit{P-Cygni} profile.
Terminal velocities of the last two lines are not included in the list provided in Table \ref{tab:velHen31475}, but that of \ion{He}{I} $\lambda$6678 is.

Thus, in our observations we can  identify several helium lines and we find two distinct profiles: pure \textit{P-Cygni} and broad emissions resembling a bump. It is important to note that \ion{He}{I} lines belong to two different families. When ionized helium recombines, the captured electron can have either parallel or anti-parallel spin relative to the spin of the electron present in the level, leading to triplet and singlet states. In the triplet state, electrons reach the metastable level ($2^3 S$) where collisional excitation can occur. We identified the bumps as transitions in the triplet state, suggesting that we could be observing a broadening due to collisional excitation. 
Furthermore, at high density the metastable level $2^3 S$ can become optically thick, affecting the observed \ion{He}{I} line profiles. In such conditions, resonant scattering and photon trapping may lead to profile modifications, such as broadening or self-absorption features. 
The lack of some lines coming from the singlet states is due to blending with prominent \ion{Fe}{II} and \ion{Ca}{II} lines.

\subsection{The O\,I lines}

Several \ion{O}{I} lines are detected throughout the spectrum, including both recombination and collisionally excited lines. Emission lines at 6300 and 6363 {\AA} are observed in 2006, each displaying a single component (this situation is different in 2024 observations, see \S \ref{sec:analysis2024}). 

\begin{figure} 
    \centering
    \includegraphics[width=0.95\linewidth]{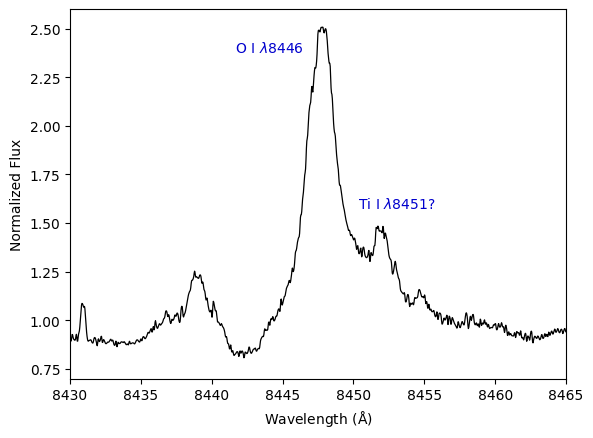}
    \caption{The \ion{O}{i} 8446 {\AA} emission observed corresponds to a triplet consisting of the lines at 8446.25 {\AA}, 8446.36 {\AA}, and 8446.76 {\AA}.}
    \label{fig:oi4}
\end{figure}

\begin{figure}
\centering
  \includegraphics[width=0.95\linewidth]{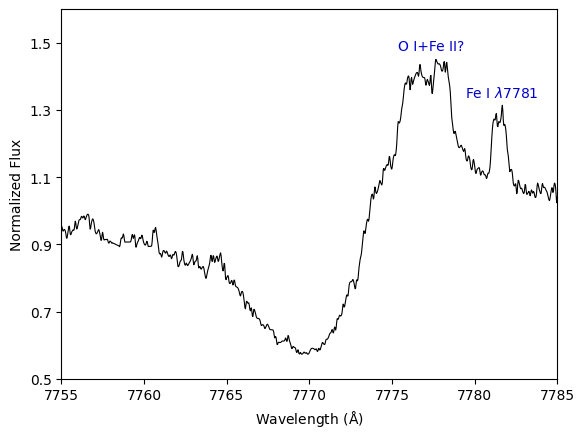}
  \caption{Absorption dip next to a series of blended lines at \ion{O}{I} $\lambda$7773.}
  \label{fig:blend}
\end{figure}

One of the most significant lines observed is \ion{O}{I} $\lambda$8446  (see Fig.  \ref{fig:oi4}). 
\citet{Riera95} noted that the presence of this line could not be explained solely by recombination processes, as this would result in the detection of a strong \ion{O}{I} $\lambda$7773 triplet. In Figure \ref{fig:blend}, a \textit{P-Cygni} profile in the $\lambda$7773 wavelength is observed, where the absorption dip displays a small bump alongside a series of blended lines. It remains unclear whether any of these features correspond to the aforementioned triplet or to another set of lines.
Two alternative mechanisms have been proposed to explain the origin of \ion{O}{I} $\lambda$8446: direct excitation from starlight and \ion{Ly}{$\beta$} fluorescence. If direct excitation were the primary mechanism, emission lines at the 7002 {\AA} doublet and at 7254 and 13164 {\AA} would be expected. In this work, neither the 7002 {\AA} doublet nor the 7254 {\AA} line are detected. Regarding fluorescence, emission lines should appear at $\lambda\lambda$ 1304 (this triplet consists of the lines at 1302.17 {\AA}, 1304.86 {\AA} and 1306.03 {\AA}) and 11287 {\AA}. While the presence of either of the IR lines has yet to be confirmed, the UV emission lines will be discussed below. It is worth noting that for AGNs, \ion{Ly}{$\beta$} fluorescence has been suggested as the primary excitation process for \ion{O}{I} 8446 {\AA} since \citet{1980Grandi}.

\subsubsection{UV spectra}

For this discussion, UV data were retrieved from the Mikulski Archive for Space Telescopes (MAST) to analyze the presence of the \ion{O}{I} 1304 {\AA} line. The goal is to explore the origin of the \ion{O}{I} 8446 {\AA} emission line.

The first UV spectroscopic observations were obtained in 1989 using the IUE satellite with the prime camera (SWP), under S. Pottasch as P.I. No prominent emission lines were detected in this spectrum, most probably due to the obscuration of the star by a dusty torus, as discussed in \citet{2003GaubaPartha}. Consequently, the \ion{O}{I} emission line at 1304 {\AA} was not observed.

More recent UV spectra were obtained in 2015 using the Space Telescope Imaging Spectrograph (STIS) aboard the HST, with X. Fang as P.I. These observations are thoroughly discussed in \citet{Xuan}.
Two gratings were used: G140L for the FUV spectrum (1150 {\AA}–1730 {\AA}) and G230L for the NUV spectrum (1570 {\AA}–3180 {\AA}). The exposure times were 2800 s for G140L and 2100 s for G230L.

\begin{figure}
    \centering
    \includegraphics[width=1\linewidth]{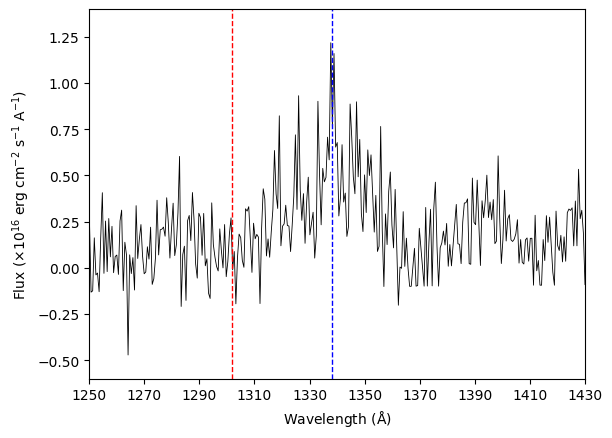}
    \caption{Zoom-in of the STIS UV spectrum obtained in 2015. The red dotted line indicates the expected position of the \ion{O}{I} 1302 {\AA} line, while the blue dotted line marks the peak of an unidentified line at $\sim$1338 {\AA}.}
    \label{fig:zoom-in}
\end{figure}

Focusing on the wavelength range of interest, Fig. \ref{fig:zoom-in} shows an emission line with a peak at $\sim$1338 {\AA}, this particular line is not discussed in the article mentioned above and its origin is not clear. On the other hand, the \ion{O}{I} triplet at $\lambda\lambda$1304, which would indicate fluorescence, does not appear. An explanation for the absence of this line has been provided by \citet{1983Oegerle}, stating that \ion{O}{I} 1304 {\AA} is not expected to be very strong relative to the local continuum. Additionally, resonance-line scattering is likely to reduce this emission below detectable limits in most Be stars. Therefore, the results obtained from the UV are not conclusive. The absence of the \ion{O}{I} 1304 {\AA} line does not invalidate the fluorescence scenario. Considering the intensities observed in the optical and the presence of an intense field of Ly {\footnotesize $\beta$} photons, fluorescence is still the most consistent dominant process to produce O {\footnotesize I} $\lambda$8446 en Hen\,3-1475.

\subsection{The Ca lines}
\begin{figure*}
\centering
\begin{subfigure}{.5\textwidth}
  \centering
  \includegraphics[width=0.95\linewidth]{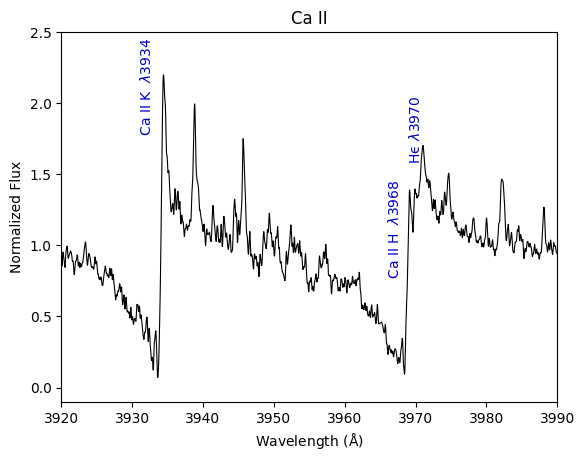}
  \caption{}
  \label{fig:CaIIlines}
\end{subfigure}%
\begin{subfigure}{.5\textwidth}
  \centering
  \includegraphics[width=0.95\linewidth]{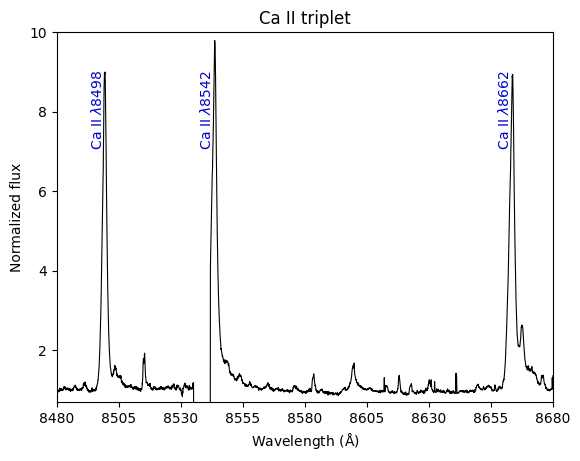}
  \caption{}
  \label{fig:calltriplet}
\end{subfigure}
\caption{(a) \textit{P-Cygni} profile in \ion{Ca}{II} H and K lines and \ion{H}{$\epsilon$}, the wing coming from \ion{Ca}{II H} and \ion{H}{$\epsilon$} blends. (b) Intense \ion{Ca}{II} lines in the IR.}
\label{fig:test}
\end{figure*}

As for the calcium lines, \ion{Ca}{II} H and K lines show very prominent \textit{P-Cygni} profiles as shown in Fig.  \ref{fig:CaIIlines} with terminal velocities up to $\sim 690$ km s$^{-1}$. In the red part of the spectrum (Fig. \ref{fig:calltriplet}) we observed the \ion{Ca}{II} triplet at 8498, 8542, and 8662 \AA, which are the strongest lines in the entire spectrum. 

\citet{Riera95} derived the optical depth in \ion{Ca}{II} 8542 {\AA} line from the ratio $\lambda\lambda\,8498$/$8542$, obtaining a value of 0.8. From this ratio, they found $\tau(8542\,\text{{\AA}})>30$, indicating an optically thick medium. In this work, we derived a ratio of 0.86 for the Ca lines; this implies a still optically thick region where \ion{Ca}{II} lines are forming. According to \citet{1988Persson} this also implies a hydrogen column density of N$_{\text{Ca}}$(H)$>8\times 10^{20}$ cm$^{-2}$. 

The [\ion{Ca}{ii}] doublet $\lambda\lambda$7291,7324 is also observed in our spectra. These lines provide information about the state of the element. Being forbidden lines, their presence indicates low density (n$_e$  about 10$^5$ - 10$^7$ cm$^{-3}$), low ionization, and a shock-heated gas. 

According to \citet{2008Sanchez}, since calcium is typically trapped in dust grains in space, the detection of [\ion{Ca}{ii}] emission lines indicates that in this case, calcium is not totally bound to the grains. This could occur because the emission originates from a dust-free region or because some process has released calcium from the dust grains, returning it to the gas phase.

\begin{table}[h]
\centering
\caption{Terminal velocity ($v_{\infty}$) found in Hen\,3-1475}
\label{tab:velHen31475}
\renewcommand{\arraystretch}{1.2}
\begin{tabular}{cccc}
\hline
\textbf{Ion} & \(\boldsymbol{\lambda_0}\) & \(\boldsymbol{\lambda_{\infty}}\) & \(\boldsymbol{v_{\infty}}\) \\ 
& \textbf{({\AA})} & \textbf{({\AA})} & \textbf{(km/s)} \\ \hline\hline
\ion{Ca}{II} K    & 3933.66 & 3925.15 & 686.95 $\pm$ 76.38 \\ \hline
H 8      & 3889.05 & 3881.25 & 639.79 $\pm$ 69.20\\
\ion{H}{$\delta$} & 4101.74 & 4090.99 & 824.33 $\pm$ 73.24\\ 
\ion{H}{$\gamma$} & 4340.47 & 4332.10 & 617.37 $\pm$ 69.20\\ 
\ion{H}{$\beta$}  & 4861.33 & 4850.49 & 707.17 $\pm$ 61.89 \\ \hline
\ion{Fe}{ii} 27   & 4128.75 & 4124.01 & 328.40 $\pm$ 72.69 \\ 
\ion{Fe}{ii} 27   & 4233.17 & 4228.36 & 378.80 $\pm$ 70.90\\ \hline
\ion{Fe}{ii} 38   & 4549.47 & 4543.89 & 406.28 $\pm$ 65.98\\ 
\ion{Fe}{ii} 38   & 4583.84 & 4578.06 & 416.54 $\pm$ 65.49\\ \hline
\ion{Fe}{II} 42   & 4923.93 & 4913.55 & 670.76 $\pm$ 61.10\\ 
\ion{Fe}{II} 42   & 5018.44 & 5008.10 & 656.22 $\pm$ 59.87\\ 
\ion{Fe}{II} 42   & 5169.03 & 5158.18 & 667.81 $\pm$ 58.12\\ \hline
\ion{Fe}{II} 48   & 5362.87 & 5354.67 & 462.70 $\pm$ 55.99\\ \hline
\ion{Fe}{II} 49   & 5276.00 & 5271.15 & 313.75 $\pm$ 56.90\\ 
\ion{Fe}{II} 49   & 5316.62 & 5309.86 & 419.52 $\pm$ 56.46\\ \hline
\ion{Fe}{II} 74   & 6149.26 & 6143.11 & 338.12 $\pm$ 48.80 \\ 
\ion{Fe}{II} 74   & 6238.39 & 6232.97 & 298.74 $\pm$ 48.10\\ 
\ion{Fe}{II} 74   & 6456.38 & 6451.32 & 273.22 $\pm$ 46.47\\ \hline
\ion{Si}{ii} 2    & 6347.09 & 6339.75 & 385.06 $\pm$ 47.10\\ 
\ion{Si}{ii} 2    & 6371.36 & 6365.37 & 329.74 $\pm$ 47.29\\ \hline
\ion{He}{i}       & 6678.10 & 6668.46 & 471.18 $\pm$ 44.95  \\ \hline
Pa 20      & 8392.41 & 8380.18 & 475.48 $\pm$ 38.63\\
Pa 19      & 8413.33 & 8401.43 & 462.55 $\pm$ 35.54 \\
Pa 17      & 8467.27 & 8456.93 & 404.48 $\pm$ 34.30\\ \hline
\end{tabular}
\end{table}



\section{BINARY NATURE} \label{sec:binary}

The possible binary nature of the central star of Hen\,3-1475 has been suggested in many studies. It is believed that this binary nature would be responsible for the formation of the jets \citep{Xuan} and the point-symmetric morphology, likely driven by the precession of a central binary system undergoing episodic mass-loss events \citep{2011Manteiga}.

Furthermore, \citet{2022GarciaSegura} argues that the ultra-fast wind of 2300 km s$^{-1}$, as determined by \citet{2001ApJ...553L.173S}, cannot be ruled out as a second jet formed by the companion star. 

Also, in \citet{khouri2025}, it is implied that the observed features, such as the bipolar shape, high-velocity molecular outflow, and shocks, fit well into the context of binary systems.

 All the above could lead us to assume a binary nature for Hen\,3-1475. However, we have not found differences in the radial velocity of the stellar emission in 2006 (39.34$\pm$0.38 km s$^{-1}$) and the stellar emission in 2024 (39.55$\pm$0.45 km s$^{-1}$) , which should be expected if the star were traveling in a close binary orbit. Therefore, we tend to rule out the possibility of a binary star, although the possibility of an orbit close to the sky plane cannot be discarded. However, the orientation of the bipolar outflow and jets set some constrains for the orientation of the binary system orbital plane.

\section{Analysis of the differences in 2006 and 2024 optical spectra} \label{sec:analysis2024}

\begin{figure}[h]
  \includegraphics[width=0.95\linewidth]{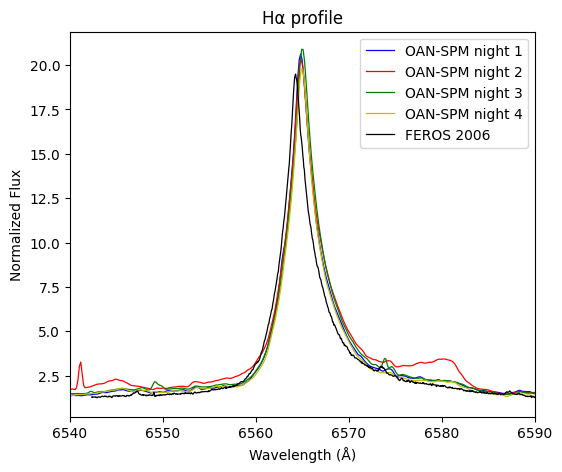}
  \caption{Superposition of H$\alpha$ emission from different nights at OAN-SPM 2024 and the FEROS 2006 observation. A difference of 94 km s$^{-1}$ between both epochs, possibly due to differences in the stellar wind, is found.} 
  \label{fig:halphas}
\end{figure}

\begin{figure}[h]
\centering
  \includegraphics[width=0.95\linewidth]{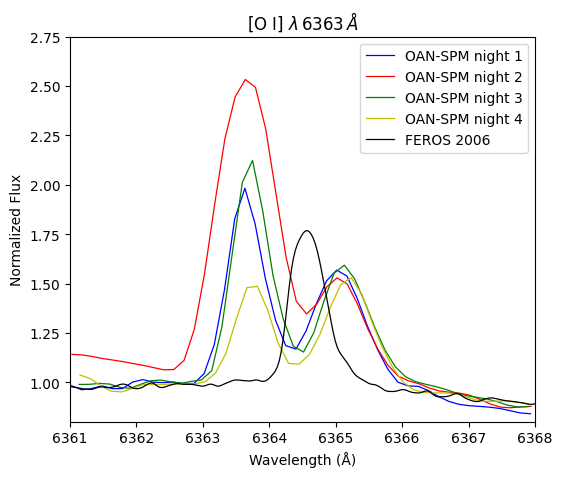}
  \caption{Two broad components of [\ion{O}{i}] at $\lambda\lambda$ 6363.9 and 6365.1 (separated by more than 60 km/s) are observed on different nights of 2024. They are compared with FEROS data, which shows 
  a single peak, corresponding to the systemic velocity, at 6364.5 \AA.}
  \label{fig:6363}
\end{figure}

In the spectra obtained at OAN-SPM, we generally observe the same features reported in the FEROS observations from nearly 20 years ago. However, one of the most notable differences appears in Figure \ref{fig:halphas}, particularly on night {\bf 2}. This figure compares the \ion{H}{$\alpha$} profiles from different nights of observation. On night {\bf 2}, a broad component adjacent to \ion{H}{$\alpha$} is observed, probably associated with the jet previously discussed in works such as \citet{Riera95}.
This emission is most prominent in the observations taken on the night of August 28 (Figure \ref{fig:halphas} in red), as it is the only set of data collected with a significantly wider slit, allowing us to capture additional zones of the nebula, facilitating the detection of wider emission and the detection of  this feature more clearly. However, traces of this broad component are also detectable in all other observations. 

As previously mentioned, the smallest slit width we used in OAN-SPM observations is a 2 arcsec one. With this width, we detected emission that could potentially originate from the jet or the shocked region, as indicated by the presence of two broad components in Fig. \ref{fig:6363}. The shown double emission  corresponds to  [\ion{O}{I}] $\lambda$6363, although it is also detected in [\ion{O}{I}] $\lambda$6300. The left component occurs at $\lambda$6363.9 and the right one at $\lambda$6365.1; therefore, they are separated by about 60 km s$^{-1}$, and it could correspond to a new jet emerging from the central zone. Such a jet is not detected in the FEROS data from 20 yr earlier. The FEROS emission lies in the middle of 2024 emissions at a wavelength corresponding with the system velocity. 
Notably, the study by \citet{Riera95}, which employed a slit width of 1.5 arcsec, did not report this shocked emission in the central spectrum either.

In \citet{2011Manteiga}, the detection of the nebular emission line [\ion{Ne}{II}] at 12.8 $\mu$m is reported, suggesting that photoionization in the central region may have already begun. However, the possibility that this line originates from shock excitation in the high-velocity outflows is not ruled out. In our work, we did not observe evidence of photoionization, as no collisional excitation lines, such as those of sulfur or nitrogen, are detected in the central region. Therefore, the detection reported by \citet{2011Manteiga}  is probably due to shocks.

\section{The knot NW1 }\label{sec:NW1}

NW1 is the brightest knot in the S-shaped string of Hen\,3-1475, and it is a zone where X-ray emission has been detected. Considering the observed size of the knot, about 0.5 arcsec, and the GAIA distance of 5.32 pc, the knot's size is approximately 1.29$\times 10^{-5}$ pc. This knot was extensively analyzed by \citet{2006Riera}, and it is interesting to investigate any differences in its characteristics more than 20 years later. 

\begin{figure}
\centering
  \includegraphics[width=0.95\linewidth]{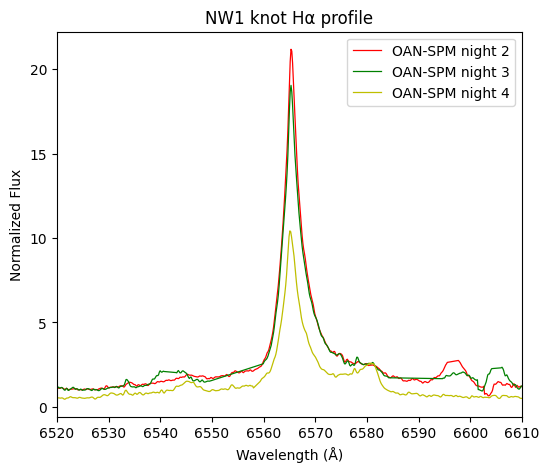}
  \caption{\ion{H}{$\alpha$} emission from observations of knot NW1 and jet emission.}
  \label{fig:halphaNW1}
\end{figure}

\begin{figure}
    \centering
    \includegraphics[width=0.95\linewidth]{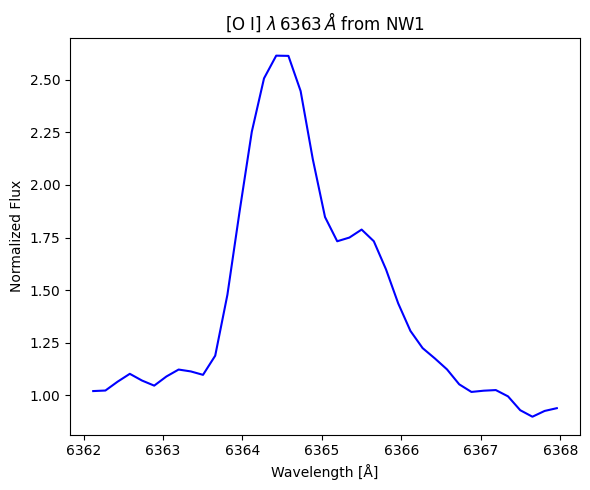}
    \caption{Double-peaked profile of the [\ion{O}{I}] 6363 {\AA} emission line originating from the knot NW1.}
    \label{fig:6363NW1}
\end{figure}

Figure \ref{fig:halphaNW1} shows a comparison of \ion{H}{$\alpha$} emission from the knot on different nights. In general, the spectra of nights 2 and 3 (slit P.A. E-W) seem to be a reflection of the emission of the star due to dust grains, as the most intense stellar lines are still seen but much more diminished in flux.   On the other hand, on night 4 (August 30, slit P.A. N-S) the continuum and \ion{H}{$\alpha$} emission are lower than on the other nights and we see that the jet appears much more marked (emission at $\lambda$6580), with more prominent emission from the shocked lines such as [\ion{S}{ii}] $\lambda\lambda$6716 and 6731 and [\ion{O}{i}]$\lambda\lambda$6300 and 6364,  all of which present a double  peak, as shown in Fig. \ref{fig:6363NW1}.  
It is important to remember that sulfur lines are not detected in the central region, contrary to observations of the knot, and previous studies have classified this type of emission as shock-excited.

A very interesting difference between previous results and ours on NW1 data is the following: In \citet{2006Riera} observations performed in 1999 with the STIS spectrograph, collisionally excited lines such as [\ion{O}{iii}] $\lambda\lambda$4959,5007, [\ion{O}{ii}] $\lambda$3727, [\ion{N}{ii}] $\lambda$6583, [\ion{S}{ii}] $\lambda\lambda$6716,6731 and other forbidden lines were well detected in knot NW1 and attributed to shocks. In our 2024 spectra, several of these lines are not observed. In particular, [\ion{O}{iii}] lines are absent. Instead, [\ion{O}{ii}] $\lambda$7325 and [\ion{S}{ii}] $\lambda\lambda$6716,6731 appear with very weak emission, while [\ion{N}{ii}] $\lambda\lambda$6548,6583 appear blue-shifted by 3 {\AA} and much weaker than in 1999. 

The above fact may be attributed to a combination of factors, including possible differences in the position of the slit, although our 2 arcsec slit width should be including the whole knot. 
These spectral differences may depend mostly on changes in the physical state of the gas. One possibility is that the shock velocity of $150-200$ km s$^{-1}$ or larger, needed to compute a model to explain the mentioned emissions  \citep{2006Riera}, has decreased to less than 50 km s$^{-1}$ in which case the temperature of the zone is not enough to excite the [\ion{O}{III}] 5007 \AA~ and other forbidden lines (A. Rodr{\'\i}guez-Gonz\'alez, private comm.). It is important to note that the shock model at velocities of $150-200$ km s$^{-1}$ proposed in the work by \citet{2006Riera} does not accurately reproduce the observations of [\ion{O}{III}] 5007 \AA~ in the knot NW1, suggesting that larger velocities are needed or that additional physical processes may be involved.

Another possible explanation is that during \citet{2006Riera}’s observations, the gas was photoionized by radiation from a hotter central star. Note that the central star could suffer from fast variations in temperature in its transitioning from post-AGB to PN central star, as it has occurred with, for instance, the central star of the young PN IC\,4997 \citep{2009ARep...53.1155K}. Later, if the stellar temperature decreased, the gas could have recombined quickly due to its high density.
The recombination timescale of an ionized region depends on the electron density $n_e$ and is estimated as $t_{\text{rec}} \sim 10^5 / n_e$  yr \citep{1989Osterbrock}. For $n_e \sim 3000$ cm$^{-3}$ \citep{2006Riera}, the gas could have recombined within a couple of decades. If the stellar temperature decreased between their observations and ours, the gas may have recombined, causing emission lines such as [\ion{O}{III}] 
$\lambda 5007$ to disappear. 

A list of the most prominent emission lines detected in knot NW1 during our observations in OAN-SPM is provided in Table \ref{tab:tab1NW1}.
The table includes the central wavelengths, normalized fluxes and their errors, and FWHM of lines.
Unfortunately, we have no observations of the knot NW1 in 2006 to compare the state of the ionization with the other epochs. 
\begin{table}
\centering
\caption{Spectral lines found in the  knot NW1 of Hen\,3-1475.}
\label{tab:tab1NW1}
\renewcommand{\arraystretch}{1.1}
\resizebox{0.5\textwidth}{!}{%
\begin{tabular}{lccccc}
\hline
\textbf{Ion} & \(\mathbf{\lambda_0}\)  & \(\mathbf{\lambda_{obs}}\) & Norm. &  $\sigma_F$ & FWHM \\
 & ({\AA)} & ({\AA)}  & Flux & & ({\AA}) \\\hline\hline
H$\beta$                 & 4861.33 & 4862.77 & 13.48 & 2.92 & 3.55 \\
Fe {\footnotesize II} 42 & 4923.93 & 4925.37 & 4.61 & 0.66 & 3.25 \\
Fe {\footnotesize II} 42 & 5018.44 & 5020.04 & 4.15 & 0.84 & 3.10 \\
Fe {\footnotesize II} 42 & 5169.03 & 5170.51 & 8.30 & 0.47 & 4.38 \\
Fe {\footnotesize II} 49 & 5197.58 & 5198.01 & 5.25 & 0.33 & 5.03 \\
Fe {\footnotesize II} 49 & 5275.99 & 5276.39 & 3.30 & 0.26 & 4.25 \\
Fe {\footnotesize II} 49 & 5316.62 & 5317.83 & 3.81 & 0.51 & 2.43 \\
Fe {\footnotesize II} 49 & 5425.26 & 5426.48 & 1.50 & 0.12 & 2.20 \\ 
Na {\footnotesize I} D   & 5889.95 & 5890.47 & 0.63 & 0.10 & 1.28 \\
Na {\footnotesize I} D   & 5895.92 & 5896.44 & 1.15 & 0.15 & 1.81 \\
{[}O {\footnotesize I}{]} & 6300.30 & 6300.64 & 10.47 & 4.07 & 1.13 \\
{[}O {\footnotesize I}{]} & 6300.30 & - & - & - & - \\
{[}O {\footnotesize I}{]} & 6363.78 & 6364.45 & 1.57  & 0.31 & 0.92 \\
{[}O {\footnotesize I}{]} & 6363.78 & 6365.57 & 0.63  & 0.22 & 0.92  \\
{[}N {\footnotesize II}{]} & 6548.05 & 6545.27 & 3.86 & 0.24 & 5.32 \\
H$\alpha$ & 6562.82 & 6565.52 & 39.35 & 7.57 & 4.87 \\
H$\alpha$ jet/ {[}N {\footnotesize II}{]}? & 6562.82 &  6580.17 & 6.50 & 0.77 & 4.33  \\
{[}S {\footnotesize II}{]} & 6715.77 & 6714.10 & 3.77 & 0.19 & 4.82 \\
{[}S {\footnotesize II}{]} & 6730.82 & 6728.54 & 3.75 & 0.30 & 3.78 \\
\hline
\end{tabular}}
\end{table}

\section{Conclusions}\label{sec:conclusion}

High-resolution spectra of the central zone of Hen\,3-1475, obtained in 2006 and 2024, have been deeply analyzed. More than 500 emission lines, including lines from \ion{H}{I}, \ion{He}{I}, \ion{Fe}{II}, [\ion{Fe}{II}], \ion{O}{I}, [\ion{O}{I}], \ion{Ca}{II}, [\ion{Ca}{II}], \ion{Si}{II}, and other ions, have been detected.  The complete list of lines with their normalized fluxes is presented in Table \ref{tab:Hen31475}.

\textit{P-Cygni} profiles are identified in many lines, indicating the presence of high-velocity stellar winds.  This type of profile is particularly observed in \ion{H}{I}, \ion{He}{I}, some \ion{Fe}{II} lines, \ion{Ca}{II} and \ion{Na}{I}  lines. The wind terminal velocity was measured for these lines, and the results are presented in Table \ref{tab:velHen31475}. The wind velocity varies from about 300 km s$^{-1}$ for the \ion{Fe}{II} lines to about 700 km s$^{-1}$ for the H Balmer lines. 

Several \ion{He}{I} lines are present in these spectra; the \ion{He}{I} $\lambda$6678 and $\lambda$7065 lines were detected, presenting \textit{P-Cygni} profiles. On the other hand, a bump not reported before is identified in \ion{He}{I} $\lambda$5876,  
$\lambda$ 4471 and two other lines. These bumps are detected in FEROS as well as in OAN-SPM observations. The origin of this emission, which spans approximately $\sim 10$ {\AA}, is unclear although wide stellar emission lines are typically observed in stars with strong massive winds.
The lines showing bumps correspond to \ion{He}{I} triplets while the singlets  exhibit only a \textit{P-Cygni} profile without broad or bumpy features. This could be attributed to the large optical depth affecting the \ion{He}{I} triplets and not the singlets \citep{2022Kwitter}.

The presence of the \ion{O}{I} $\lambda$8446 line is analyzed, as the origin of this triplet has been widely debated. It is often attributed to Ly~$\beta$ fluorescence emission, although direct excitation by stellar radiation may also contribute. This mechanism implies the presence of additional emission lines at $\lambda$1304 in the UV and $\lambda$11287 in the NIR. We attempted to detect the UV line, as no NIR observations extending to $\sim$12000~{\AA} are available. No emission is observed at $\lambda$1304 in the UV spectrum; however, this absence may be due to the expected low intensity of the line relative to the continuum or to resonance-line scattering, which can reduce the emission below detectable limits in most Be stars, as suggested by several authors \citep{1983Oegerle, 2013Marziani}. Therefore, Ly~$\beta$ fluorescence remains the most likely explanation for the observed \ion{O}{I} $\lambda$8446 emission.

The possible binary nature of the central star is also discussed. While many studies have proposed the central star of Hen\,3-1475 as a binary system, this has never been confirmed. In Hen\,3-1475, the presence of a jet and a nebula with an S-shaped string of point-symmetric knots is known, and such structures have been theorized to originate from binary systems. 
However, the central star of Hen\,3-1475 is  visible and no secondary object has been detected so far. Besides, in this work we found that the radial velocity of Hen\,3-1475 stellar lines  is the same in 2006 and 2024 observations, casting doubts on the possibility of a star traveling in a close binary orbit.

The systemic velocity derived for OAN-SPM observations is v$_{\text{LSR}}\approx 39.56 \, \pm$ 0.45 km s$^{-1}$. This is consistent with our FEROS data, considering the errors in measurements. It is also consistent with the systemic velocity reported by \citet{2003Riera} and references therein.

The analysis of FEROS data obtained in 2006 and OAN-SPM data obtained in  2024  indicates some significant changes in the spectral features over nearly two decades. Variations in slit width and resolution during 2024 observations allowed the detection of additional features, such as shock-related emissions. In 2024, we observe two components in the [\ion{O}{I}] $\lambda\lambda$ 6300, 6363 lines in the central zone, not observed in 2006. These lines are separated by about 60 km s$^{-1}$ that most likely come from shock excitation and could be due to a new jet emerging from the central zone. Also, a red broad component adjacent to \ion{H}{$\alpha$} is present at 6580 {\AA} (see Figure \ref{fig:halphaNW1}), that is possibly associated with the object's jets. More specifically, we consider it as \ion{H}{$\alpha$} emission originating from a jet at more than 700 km s$^{-1}$.

In 2024, the bright knot NW1, located in the S-shape string of Hen\,3-1475, at 2.8 arcsec from the center, was observed. Interestingly, we found that emission lines such as [\ion{O}{III}] $\lambda$5007, which were reported  by \citet{2006Riera} and attributed to shocks, were not detected, showing that the velocity of the shocks has possibly diminished in these 20 years.

In both 2006 and 2024 data, the absence of typical photoionized nebular lines such as [\ion{O}{III}] $\lambda$5007 implies that the present central star's effective temperature is lower than 30,000 K. However, the star is changing fast, and it should be kept under observation to determine its evolution.

\section*{Acknowledgements}
This study received partial support from the Dirección General de Asuntos del Personal Académico-Programa de Apoyo a Proyectos de Investigación e Innovación Tecnológica (DGAPA-PAPIIT) under project IN\,111423. A.V.B.-S. acknowledges a scholarship from CONAHCyT, México. We are grateful to Dr. Christophe Morisset  and Dr. Diego B. Hern\'andez-Ju\'arez for their detailed review and insightful suggestions, which significantly improved this work.

This research is based on observations conducted at the Observatorio Astronómico Nacional in San Pedro Mártir (OAN-SPM), Baja California, México. We thank the support staff at OAN-SPM for their assistance in facilitating and carrying out our observations.

This work is also based on observations collected at the European Southern Observatory under program 077.D-0478(A) at La Silla, Chile.

This work makes use of IRAF, which was distributed by the National Optical Astronomy Observatory (NOAO) and operated by the Association of Universities for Research in Astronomy (AURA) under a cooperative agreement with the National Science Foundation (NSF).

\begin{appendices}
 \label{sec:ap-A}
 \onecolumn
 \section{Line identification  in the FEROS spectrum}
\begin{figure*}[htp]
\centering
  \includegraphics[width=1\linewidth]{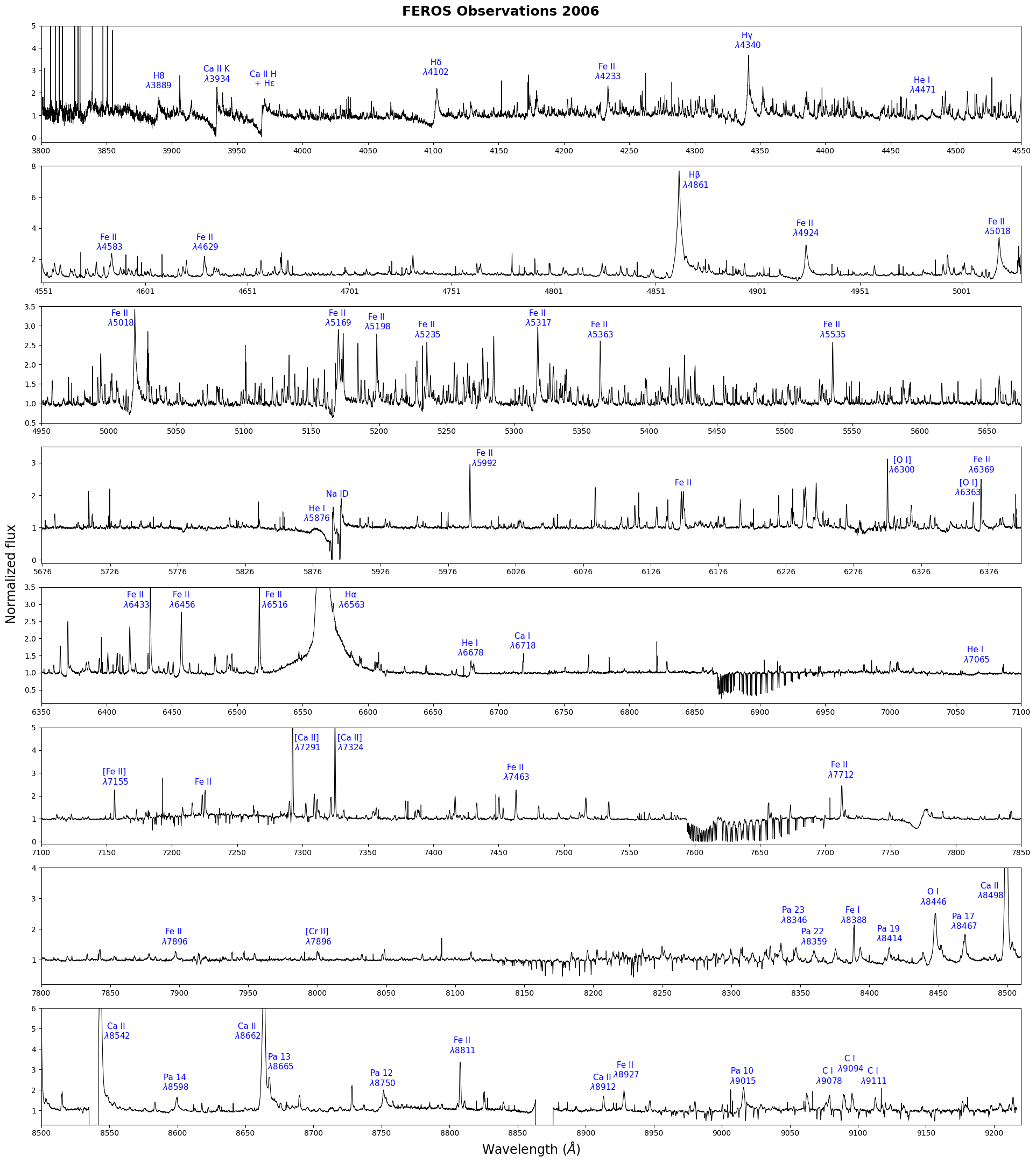}
  \caption{FEROS spectrum obtained in 2006, in the range from 3800 to 9200 \AA, normalized to the continuum. The strongest lines have been marked.}
  \label{fig:FEROS}
\end{figure*}

 In Fig. \ref{fig:FEROS} the spectrum obtained with FEROS in 2006, in the range from 3800 to 9200 \AA, normalized to the continuum, is presented, and the strongest lines have been marked.
  \newpage
\section{Second Appendix}
\label{sec:ap-B}
\onecolumn
In Table \ref{tab:Hen31475} the lines detected in the FEROS spectrum are listed. {\bf The list includes} more than 500 lines.
The emitting ion, the {\bf rest wavelength}, the observed wavelength, the normalized flux and its error ($\sigma$F), the full width at half maximum (FWHM),  
the observed profile (\textit{P-Cygni} or not), the type of line (stellar or nebular), and the systemic velocity are presented. The lines with a quotation mark have an uncertain identification.

{\renewcommand{\arraystretch}{1.37}
{\fontsize{8.6}{10}\selectfont
\begin{longtable}{P{2cm} P{1.2cm} P{1.2cm} P{1.2cm} P{0.6cm} P{1.2cm} P{1.4cm} P{1.3cm} P{1.2cm}}
\caption{Spectral lines in the central region of Hen 3-1475.} \label{tab:Hen31475} \\ \hline
\textbf{Ion} & $\lambda_0$ (\AA) & $\lambda_{obs}$ (\AA) & Norm Flux & $\sigma_F$ & FWHM & Profile & Type & $v_{sys}$ (km/s) \\ \hline \hline
\endfirsthead
\multicolumn{9}{c}{{\bfseries\footnotesize{Continue Table \ref{tab:Hen31475}}}} \\ \hline
\textbf{Ion} & $\lambda_0$ (\AA) & $\lambda_{obs}$ (\AA) &  Norm Flux & $\sigma_F$ & FWHM & Profile & Type & $v_{sys}$  (km/s) \\ \hline \hline
\endhead
\hline \multicolumn{9}{r}{{Continue in next page}} \\ \hline
\endfoot
\hline \hline
\endlastfoot
H8                             & 3889.05 & 3889.56 & 3.25              & 0.19       & 2.65 &\textit{P-Cygni} & Stellar & -        \\
Si \scriptsize{I} (3)          & 3905.52 & 3906.06 & 0.98              & 0.26       & 0.58 &                 &         & 41.25        \\
Fe \scriptsize{I} 567?         & 3914.27 & 3914.79 & 0.49              & 0.07       & 0.74 &\textit{P-Cygni} & Stellar & -        \\
Ca \scriptsize{II} K           & 3933.66 & 3934.20 & 0.92              & 0.16       & 1.08 &\textit{P-Cygni} & Stellar & -        \\
Fe \scriptsize{II} 3           & 3938.29 & 3938.83 & 0.69              & 0.10       & 0.84 &                 & Stellar & 40.90        \\
Fe \scriptsize{II} 3          & 3945.21 & 3945.71 & 0.46              & 0.09       & 0.61 &                 & Stellar & 37.94        \\
Ca \scriptsize{II} H           & 3968.47 & 3969.06 & -                 & -          & -    &\textit{P-Cygni} & Stellar & -        \\
H$\epsilon $                   & 3970.07 & 3970.55 & -                 & -          & -    & \textit{P-Cygni}& Stellar & -        \\
\hline
\end{longtable}}}

\end{appendices}
\bibliography{example}

 \end{document}